\documentclass[12pt,preprint]{aastex}

\slugcomment{31 Dec, 2012}
\shorttitle{SUBARU/HDS STUDY OF HE 1015$-$2050:
SPECTRAL EVIDENCE OF R CORONAE BOREALIS LIGHT
DECLINE }
\shortauthors{Aruna Goswami et al.}
\begin{document}
\title{SUBARU/HDS STUDY OF HE 1015$-$2050:
SPECTRAL EVIDENCE OF R CORONAE BOREALIS LIGHT
DECLINE  }
\author{Aruna Goswami\altaffilmark{1} and  Wako Aoki\altaffilmark{2} }
\affil{$^{1}$Indian Institute of Astrophysics, Bangalore 560034, India;
  aruna@iiap.res.in \\
$^{2}$National Astronomical Observatory, Mitaka, Tokyo 181-8588, Japan
}

\begin{abstract}
Hydrogen deficiency and a sudden optical light decline by about 6-8 mag 
are two principal characteristics of R Coronae Borealis (RCB) stars. The high 
latitude carbon star HE~1015$-$2050 was identified as a hydrogen-deficient 
carbon star  from low-resolution spectroscopy. 
 Photometric data  of the Catalina Real-Time Transient Survey gathered
between  2006 February and  2012 May indicate that the object exhibits no 
variability. However, a 
high-resolution ( R ${\sim}$ 50,000) optical spectrum of this object 
obtained with the  8.2m Subaru telescope using High Dispersion Spectrograph
 on the  2012 January 13  offers  sufficient spectral evidences for the 
object being a 
cool HdC star of RCB type undergoing  light decline. In contrast to 
the Na I D broad absorption features, seen in the low-resolution 
spectra  on several occasions, the high-resolution spectrum exhibits  
Na I D$_{2}$ and D$_{1}$ features in emission. A few  emission lines 
due to Mg I, Sc II, Ti I, Ti II,  Fe II and Ba I are also observed 
in the spectrum of this object for the first time. Such emission 
features combined with  neutral and singly ionized lines of Ca, Ti, 
Fe,  etc.,  in absorption  are reportedly seen  in RCBs spectra in the 
early stage of decline or during the recovery to maximum. Further, the 
light decline of RCBs  is  ascribed to the formation of a cloud of 
soot that obscures the visible photosphere. Presence of such
circumstellar material is evident from the  polarimetric observations 
with an  estimated V-band percentage polarization of  ${\sim}$1.7\% 
for  this object. 

\end{abstract}

\section{INTRODUCTION}

Hydrogen-deficient carbon (HdC) stars and   R Coronae Borealis
(RCB) type stars  form a rare class of carbon-rich supergiants
(\citet{Clayton96}, \citet{Clayton12} for a general review).
The deep minima and infrared excesses that are characteristics
of RCB stars are absent in HdC stars (\citet{Warner67}; \citet{Feast73}; 
\citet{Feast97}). These stars have spectra similar to F-G supergiants, 
but show only weak absorption features of hydrogen while molecular 
bands of carbon are strong (in cool RCBs and HdCs). The most promising
scenario for the origin of these objects is that they are 
post-asymptotic giant branch (post-AGB) stars, which are low- to 
intermediate-mass stars at the evolutionary state from AGB to white 
dwarfs, particularly objects which experience a final helium flash 
(\citet{Iben96}). However, many other scenarios, including a merger 
of two white dwarfs, have also been proposed  (\citet{Webbink84}; 
\citet{Renzini90}; \citet{Staff12}).

Chemical composition studies of RCB stars suggest the  presence of
material in their atmosphere is being  exposed to s-processing.
\citet{Asplund00} find that all RCB stars show some enhancement 
of s-process elements relative to Fe ([Y/Fe] = 0.8; [Ba/Fe] = 0.4). 
Extraordinary overabundances of Y and Ba in U~Aqr are also reported 
by \citet{Vanture99} with estimates of  [Y, Zr/Fe] ${\ge}$ +3.0
and [Ba/Fe] = 2.1. Light neutron-capture elements like Sr, Y, and Zr 
are usually attributed to the weak component of the s-process, which 
is the neutron-capture process with relatively long time scale and 
small neutron exposure (\citet{Kappeler11}). Massive stars in 
He- or C-burning phase are suggested as the astrophysical site of the 
weak s-process, but there is no direct observational evidence to
support this scenario. 
 
From low-resolution spectroscopic analysis of a large number of faint 
high latitude carbon stars of Hamburg/ESO survey, \citet{Goswami10} 
have identified HE~1015$-$2050 to be an  HdC star. The object has an 
IR excess, the observed  magnitudes detected by NASA's {\it Wide-field 
Infrared Survey Explorer} (WISE) are 14.663 at  3.4 and 4.6 ${\mu}$m 
bands, 12.887 at 12 ${\mu}$m and 9.195 at 22 ${\mu}$m band. The object is 
also monitored by the Catalina Real-Time Transient Survey (CRTS); 
light-curve data obtained between  2006 February and  2012  May indicate 
that the object exhibits no variability at the level of measurement
uncertainties  (Andrew Drake, private communication 2012).
 In this Letter, we discuss its high-resolution spectrum  and  confirm 
its hydrogen deficiency. Further, we find that the spectrum displays  
features that are reminiscent of RCBs at early decline or near 
recovery to maximum. The spectrum is characterized by strong features 
of  Sr and Y. There are no obvious stronger features of heavier 
elements (i.e., Mo, Ba) indicating that this object can be a source 
of weak s-process.

\section{OBSERVATIONS AND DATA REDUCTION}

The high-resolution spectroscopic observations of HE~1015$-$2050 was 
carried out with the High Dispersion Spectrograph (HDS) of the 8.2m Subaru
Telescope (\citet{Noguchi02}) on  2012 January 13. The  spectrum was
taken with a 30 minute exposure at  a resolving power of $R \sim$
50,000. The observed bandpass ran from about $4010$ \AA\ to
$6800$ \AA\, with a gap of about $75$ \AA\, from $5355$ \AA\ to
$5430$ \AA\, due to the physical spacing of the CCD detectors. 
The low-resolution (R ${\sim}$ 1300) spectrum obtained with 2m HCT
at IAO, Hanle is taken from (\citet{Goswami10}). The data were 
reduced, in the standard fashion, using IRAF\footnote{IRAF is 
distributed by the National Optical Astronomical Observatory, which 
is operated by the Association for Universities for Research in 
Astronomy, Inc., under contract to the National Science Foundation.} 
spectroscopic data reduction package.

\section{RADIAL VELOCITY}
A selection of about 30  absorption lines are used to determine  the 
radial velocity of the object HE~1015$-$2050.  Velocities are derived  
from the central cores of  lines. Mean  Heliocentric radial velocity is 
estimated to be ${\sim}$ 18 ${\pm}$ 1.5 km s$^{-1}$. Radial velocity 
survey of RCB stars by \citet{Lawson96} records a typical 
peak-to-peak variations of  10-20 km s$^{-1}$. The velocities estimated 
from the emission features of  H$_{\alpha}$, Na I D$_{2}$ and D$_{1}$, 
Sc II ${\lambda}$5684.19  are,  respectively, 17.1, 19.3, 19.6, and 
21.8 km s$^{-1}$, and  are similar to the velocity derived from the 
photospheric absorption lines. There is, however,  evidence that 
the narrow emission  lines are blueshifted relative to the 
pre-decline absorption lines  (e.g, R CrB, \citet{Cottrell90}).
 
\section{ SPECTRAL CHARACTERISTICS}

During deep declines of RCBs (e.g., R CrB and RY Sgr)  the forming dust 
cloud extinguishes the photospheric light and a rich narrow-line emission 
spectrum appears consisting of neutral and singly ionized metals 
(\citet{Payne63}; \citet{Alexander72}; \citet{Clayton96}). Most of 
these lines referred as E1 (\citet{Alexander72}) are short-lived that  
fade within two or three weeks and are replaced by a broad-line spectrum. 
Some of the early-decline narrow emission lines remain strong for an 
extended period of time. These lines referred as E2  are mostly low 
excitation lines, and primarily multiples of Sc II and Ti II. The 
Sc II ${\lambda}$4246 line is one of the strongest lines in the E2 spectrum 
of RCB stars in decline  (\citet{Rao99}). The spectrum of 
HE~1015$-$2050 shows narrow-line emission features as well as broad 
absorption features, leading us to believe that the spectrum was 
acquired  when the object was obscured by forming dust. We put forward 
our arguments supporting this idea in the following sections.

 \subsection {Spectral Characteristics: Hydrogen Deficiency}
In Figure 1, we  show a comparison of  the spectrum of  HE 1015$-$2050  
with the spectrum of HD~26, a hydrogen normal CH star, in the wavelength 
region 4310$-$4330 \AA\,. The features due to G band of CH, seen  in deep 
absorption  in  HD~26, are distinctly shallow in  HE~1015$-$2050. We 
have also compared the H$_{\alpha}$ feature in  HE~1015$-$2050 with its 
counterpart in HD~26 (Figure 1, right panel). While H$_{\alpha}$  
appears as a strong  absorption feature in  the spectrum of HD~26, in 
the spectrum of HE~1015$-$2050 we note an emission feature at the 
absorption core. The velocity estimated from the H$_{\alpha}$  emission 
feature is  ${\sim}$+17.1 ${\pm}$ 1.5 km s$^{-1}$, very similar to the    
velocity derived from the   photospheric absorption  lines.

Marginal detection or non-detectability of H$_{\beta}$ and H$_{\gamma}$ 
features could also be an evidence for hydrogen deficiency. Weak and 
shallow features of  H$_{\beta}$ (Figure 5) and H$_{\gamma}$  in the 
spectrum of HE~1015$-$2050 are  seen with marginally detected emissions 
at the  absorption cores. The Balmer lines, which are typically very 
weak  in RCB stars, are seen going  into emission  in a few cases 
(i.e.,  V~854 Cen, \citet{Rao93}; \citet{Clayton93}). 
Strong Balmer lines and  features of CH bands are seen in the 
spectrum of V~854 Cen (\citet{Kilkenny89}; \citet{Lawson89}).

\subsection{Spectral Characteristics: Carbon Molecular Bands}
 
The spectrum of HE~1015$-$2050 contains  lines of  neutral atomic 
carbon, and  lines of blue-degraded Swan system  of C$_{2}$ bands. 
The red isotopic bands involving  $^{13}$C are not normally present 
in RCB stars (Lloyd Evans et al. 1991).  As expected, the (1,0) 
$^{13}$C$^{12}$C $\lambda$4744 band  is    much weaker than the 
(1,0) $^{12}$C$^{12}$C $\lambda$4737 band. Among the C$_{2}$ molecular 
bands, $^{12}$C$^{12}$C ${\lambda}$6122, and 
$^{12}$C$^{12}$C ${\lambda}$5635 are weak but distinctly seen in 
absorption  with redshifted bandheads. Bands of 
$^{12}$C$^{13}$C ${\lambda}$6100  and $^{12}$C$^{13}$C ${\lambda}$6168 
are not detected. 

As in the case of normal RCBs the CN bands of  (4,0) ${\lambda}$6206,
(6,1) ${\lambda}$5730, and  (7,1)  ${\lambda}$5239  are weak or absent  
in the  spectrum of HE~1015$-$2050. Strong blue-degraded CN band, (0,1) 
${\lambda}$4216  is however seen in RCBs. In the spectrum  of  
HE~1015$-$2050, the strong  SrII feature that appears  at the 
location of 4215.2 \AA\,  is likely to  have been affected by 
contributions from this band. Many C I absorption lines are reportedly  
observed in RCBs in decline  (e.g., RY~Sgr)  which fill in but 
never go into emission (Alexander et al. 1972). In some declines, 
the Swan bands of C$_{2}$ are seen in emission
(\citet{Payne63}; \citet{Whitney92}; \citet{Rao93}). 

\subsection{Spectral Characteristics: Emission Features}
{\it Na I D$_{2}$ and D$_{1}$ features.}
In Figure 2, we show the Na I D features  of HE~1015$-$2050. In the 
low resolution (${\lambda}/{\delta\lambda}$ ${\sim}$ 1300) spectrum,
Na I D$_{2}$ and D$_{1}$  features are not resolved  and appear in 
absorption (Figure 2, lower panel). These features are clearly 
resolved  in the high resolution 
(${\lambda}/{\delta\lambda}$ ${\sim}$ 50,000) spectrum and appear in 
emission (Figure 2, upper panel). Emission features of Na I D$_{1}$ 
and  D$_{2}$ are reportedly seen in decline spectra of  RCBs; at 
maximum these features appear in absorption. The appearance of 
Na I D$_{2}$ and D$_{1}$ in emission in the high-resolution spectrum 
of HE~1015$-$2050 is a clear indication  that the spectrum is 
acquired  at  its  early decline or near recovery to maximum.
In the R CrB decline spectra, these two features in emission show 
a large variation throughout the decline  (Rao et al. 1999).

{\it Forbidden  oxygen  lines.}
Sharp emission features of the forbidden lines of [OI] 5577, 6300 and 
6363 \AA\, are identified in the spectrum of HE~1015$-$2050.  The 
estimated flux ratio  ([F(6300) +F(6363)]/F(5577)) is  ${\sim}$ 0.29. 
\citet{Rao99}, have reported a value of ${\sim}$18 for this flux 
ratio throughout the 1995-1996 decline of R CrB.

{\it  Atomic lines: emission features of Sc, Mg, Ti, Fe, Ba. }  
We have made a weak detection of the line Sc II ${\lambda}$4246.8 in 
emission in the spectrum of HE~1015$-$2050. This line is one of the 
strongest lines in the "E2" spectrum of RCB stars in decline (e.g., 
\citet{Rao99}). Sc II  line at 5657.88 \AA\, also  appears in 
emission. As shown in Figure 3, an emission feature at the absorption 
core of Mg I at 5167.3 \AA\, is seen in its spectrum. A few other 
emission features such as, Mg I 5183.6 \AA\,, Ti I  5173.7 \AA\,, 
Ti II 5185.9 \AA\, and  Fe II 5197.5 \AA\, are also noticed in the 
spectrum (Figure 3). As shown  in Figure 4, lines of  
Sc II ${\lambda}$5684.2 and  Ba I ${\lambda}$6498.7  are clearly 
seen  in emission. Emission lines of  Sc II ${\lambda}$5684, 
Fe II ${\lambda}$6247, Ba II ${\lambda}$6497, Fe II ${\lambda}$5362, 
Fe I ${\lambda}$${\lambda}$5405, 5371, 5447, 
Ti II ${\lambda}$${\lambda}$5490, 5492, Si I ${\lambda}$4102, 
Zr II ${\lambda}$4096  are reportedly seen in the 1995 October 18 
decline spectrum of R CrB with red absorption components 
(Rao et al. 1999). Some unidentified emission features are known 
to exist in RCB decline  spectra (\citet{Whitney92}; \citet{Asplund95});
the spectrum of HE~1015$-$2050  also displays a number of emission 
features that remain unidentified.

\subsection{Spectral Characteristics: Absorption Spectrum}

{\it Helium. } 
Among the four lines from the triplet series of neutral helium 
seen in R CrB's decline spectrum (Rao et al. 1999), except for
5876 \AA\, the other three 3889 \AA\,,  7065 \AA\,, and 10830 \AA\, 
are out of the wavelength region considered here. However, 
He I lines of  D3  5876 \AA\, are not detected either in 
emission or absorption in the  spectrum of HE~1015$-$2050. This 
line is observed in broad emission in R CrB's decline spectrum 
(\citet{Rao99}). It is  possible that this line has also been 
filled-in by emission but has not risen above the continuum.
Other He I triplets at 4471 \AA\, and  4713 \AA\, and  singlet 
lines of He I  at  4922 \AA\, and  5015 \AA\,  are  not detected 
in the spectrum of HE~1015$-$2050. 

{\it Lithium. }  
Features due to Li are not detected in the spectrum of this object.
Only four  RCB stars (e.g., UW Cen, R CrB, RZ Nor, and SU Tau) are known to 
be  Li-rich with Li abundance ranging from 2.6 to 3.5 (\citet{Asplund00}).
\citet{Pollard94} reported  the presence of Li in two of the LMC R CrB 
stars. \citet{Vanture99}  mention a tentative identification of the 
Li I 6708 \AA\, line in U~Aqr.  Z UMi, a cool RCB star  also shows 
lithium ( \citet{Kipper06}). In Figure 5, the location of Li I 
at 6708 \AA\, in the spectrum of HE~1015$-$2050 is shown,  compared with 
its counterpart in Z~Umi,  where this feature is seen strongly. The 
spectrum of Z~Umi is taken from  \citet{Goswami97}.

{\it Atomic lines: Na, Mg, Si, Ca, Ti, V, Cr, Mn, Fe, Co, Ni. }
Many neutral and singly ionized  lines of  these elements  are  
detected in the spectrum of  HE~1015$-$2050. Some of these lines show 
filled-in emission. For instance, Ca I line at 4226.7 \AA\, and Fe I  
line at 4222.2 \AA\,  detected in absorption show filled-in emission 
at the absorption cores, whereas NaI line at 5688.2 \AA\, and Ca I 
line at 4094.9 \AA\, also detected in absorption do not show any 
filled-in emission. Several neutral and singly ionized lines of 
FeI are also detected. Lines due to V, Cr, Mn, Co, and Ni are also 
seen in the spectrum in blends with contributions from other atomic 
lines.  Cr I line at 5674.1 \AA\, is clearly detected in absorption. 
A few  absorption features due to Fe I and Ti I are shown in Figure 4 
along with a Mg I absorption feature  at 5172.7 \AA\,.

{\it Atomic lines: Sr, Y, Mo, Ba, Ce. }
 The spectrum of HE~1015$-$2050 is found to exhibit strong features  
of Strontium.  Sr II ${\lambda}$4077.7 is detected  as a strong 
absorption feature that seems to show  filled-in emission. Sr II line 
at 4215.5 \AA\, is also detected as a strong feature in absorption. 
This  feature is likely to have  been affected  by contributions  
due to strong blue-degraded (0,1) CN 4216 band. These two Sr 
features are shown in Figure 5. Although detected  clearly,  Ba II 
at 4554 \AA\, (Figure 5) and 6496 \AA\, (Figure 4) are obviously not 
stronger than Sr lines.  While Ba II line at 4554.036 \AA\, 
is observed with an equivalent  width of  131m \AA\, we  have made 
a marginal detection of Ba II feature at 6496.9 \AA\,  in absorption.
Lines due to  Y II and Ce II are also detected in the spectrum  in 
blends with contribution from other atomic lines.  We have made a 
weak detection of the Mo I line at 5689.2 \AA\, in absorption 
(Figure 5). 

\subsection{ Circumstellar Environment:  Polarimetric Evidence}
The photometric light variability of RCBs  has been ascribed to 
formation of a cloud soot that obscures the visible photosphere 
(\citet{Loreta34}; \citet{OKeefe39}).  Presence of  circumstellar material
is supported by our polarimetric observations of  2012 April 21.
The  estimated  V-band percentage polarization for this object is  
quite significant with a value of  ${\sim}$1.7\% 
(\citet{Goswami13}, A. Goswami et al., in preparation for 
details). The interstellar polarization is presumed to be negligible 
because of the object's  high Galactic latitude. Information on 
polarization estimates measured  in RCB declines are limited to only 
a few objects (\citet{Serkowski69}; \citet{Stanford88}; 
\citet{Whitney92}). \citet{Coyne73} found large polarization 
variations associated  with a brightness minimum of R CrB in 1972; 
during a decline of 7 mag, the polarization in B band was 
found to vary  from 0.29\% to 3.29\%  over three  weeks. A number of 
studies have confirmed  that the dust causing the declines  is 
carbon rich (\citet{Holm82}; \citet{Hecht84}; \citet{Wright89}; 
\citet{Hecht91}). 

\section { CONCLUDING REMARKS}

The high-resolution spectrum  of the HdC star HE~1015$-$2050, acquired 
on  2012 January 13, shows features that are reminiscent of RCBs at early 
decline or near recovery to maximum.  Optical light decline in RCBs 
is attributed  to the formation of clouds of carbon dust (O'Keefe 1939). 
Presence of circumstellar material is supported by our polarimetric 
observation.

Appearance of a narrow-line emission spectrum   consisting of lines 
of neutral and singly ionized metals, and a few broad lines including 
Ca II H and K, the Na I D lines is a common characteristics of RCBs 
in decline (\citet{Payne63}; \citet{Alexander72}; \citet{Clayton96}; 
\citet{Rao99}). With emission lines of Na I D$_{1}$ and Na I D$_{2}$, 
Sc II, Mg I etc. the spectrum of HE~1015$-$2050 seems to show  
characteristics of an  RCB  decline spectrum.

The non-detection of  Li features  may be used as a  clue for the 
production mechanism of this object. Among the two most promising 
scenarios for the origin of RCBs Li enhancements are consistent with 
the FF models as in the case of Sakurai's object (\citet{Lambert86}, 
\citet{Asplund98}).  However,  a high carbon isotopic ratios as 
well as enhancement of $^{18}$O  with no Li production are expected 
from WD merger scenario. From these, the latter is a more likely 
production scenario for this object; however, knowledge of the 
carbon and oxygen isotopic ratios would be necessary to support this.

The spectrum shows strong features of Sr but no  obvious  strong 
features of heavier elements  ( i.e.,  Mo, Ba). It has been  suggested 
that the rapid proton-capture (rp)  process that  occurs  during  
the merger of a white dwarf and a neutron star is expected to 
produce a large excess of Mo (Z = 42) compared to Sr and Y, with no 
excess of barium and other heavy elements  (\citet{Cannon93}). It is
therefore unlikely that the rp-process is responsible for the 
surface composition of this object. As Thorne-Z\.ytkow objects 
are suggested to reveal the products of rp-process in their 
surface composition, HE~1015$-$2050 does not form a good candidate 
for a Thorne-Z\.ytkow object.

Strong spectral features of Sr and weak detection of Mo and  Ba 
indicate that its atmosphere is enriched with material resulting 
from weak s-process. Several possibilities exist for giving rise 
to  such   atmospheres. The object may have been  formed out of
material which is  already enriched with s-process material.
Alternatively, the object may have been in a binary system with a 
companion that had undergone unusual s-process enrichment and had
transferred mass to this object. However, the binary status of 
the object cannot be confirmed at present, and also, except for DY~Cen 
(\citet{Rao12}) none of the RCB and HdC stars  is known to be 
binary (\citet{Clayton96}). Another possibility  concerns the fact that
the spectrum does not show enhancement in heavy s-process elements, 
and hence, it seems necessary for the s-process to produce only 
light s-process material. This requires a low neutron density. 
It was shown by \citet{Smith05} that assuming all other parameters
same, higher neutron densities  produce larger amounts of heavier
elements than of lighter ones. As suggested for U~Aqr 
(\citet{Bond79}) HE~1015$-$2050 could also be an  He-C core of 
an evolved star of near-solar initial mass  that ejected its H-rich 
envelope at the He core flesh.  A single neutron exposure occurred 
at the flash, resulting in a brief neutron irradiation producing 
more of  the s-process elements. Detailed chemical composition 
studies based on high-resolution spectra taken at maximum would 
be worthwhile to unearth the production mechanism.

{\it Acknowledgment}\\
We thank  the referee Geoff Clayton for his many useful suggestions.
This work made use of the SIMBAD astronomical database, operated at CDS, 
Strasbourg, France, and the NASA ADS, USA.\\ 
\eject

\begin{figure*}
\centering
\includegraphics[angle=0,height=13cm,width=15cm]{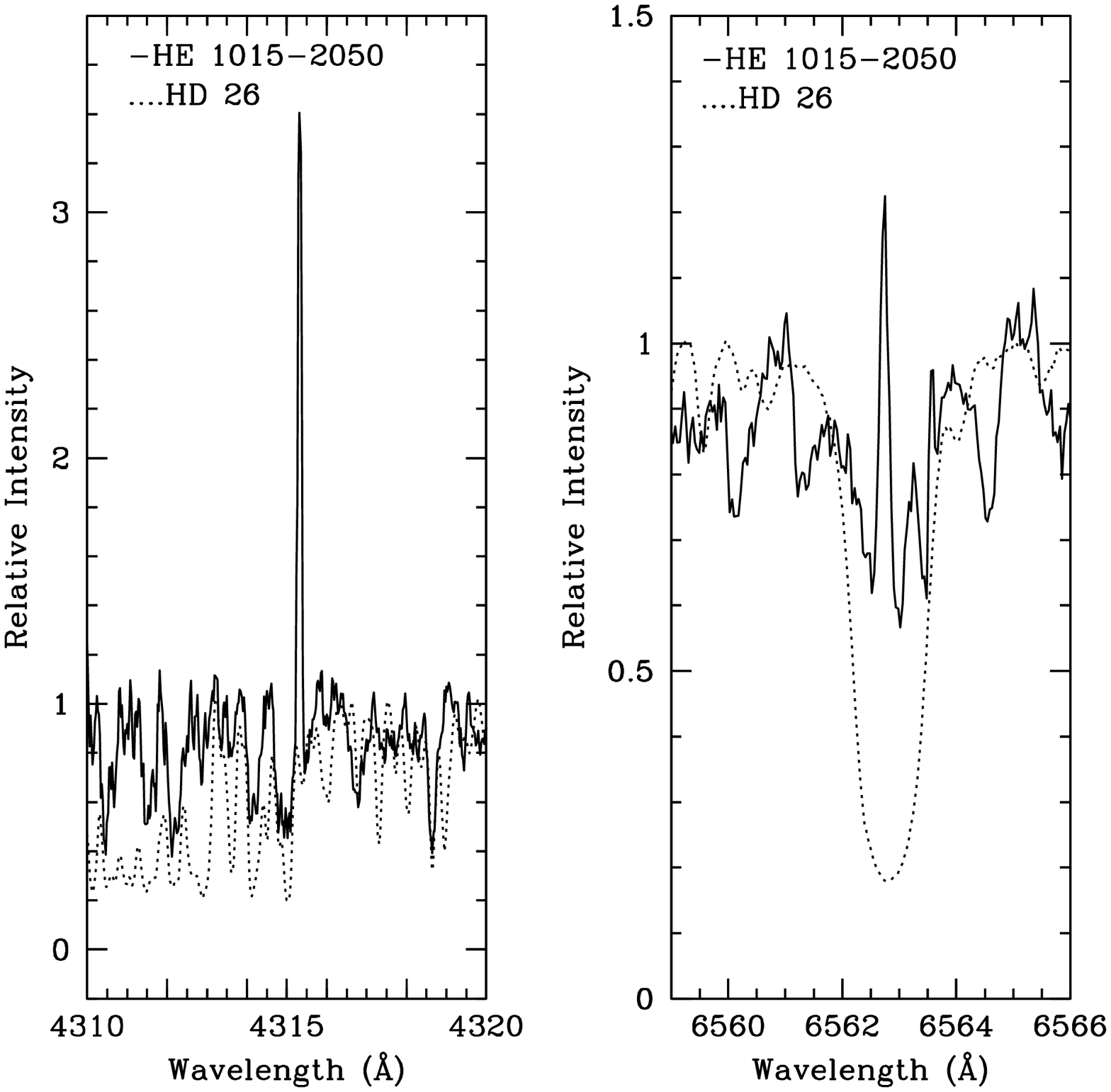}
\caption
 { Left panel: a comparison between the spectra of  HE~1015$-$2050 and 
HD~26, 
a well-known CH star in the wavelength region 4310-4320 \AA\,. Lines due to
the G band of CH around 4315\AA\, are seen in deep absorption 
in HD~26 and are much weaker in  HE~1015$-$2050.
Right panel:   while H$_{\alpha}$   appears 
as a strong absorption feature in HD~26,  an emission feature at the
absorption core is noticed in the spectrum of HE~1015$-$2050. 
\label{fig1}
} 
\end{figure*}

\begin{figure*}
\centering
\includegraphics[angle=0,height=13cm,width=15cm]{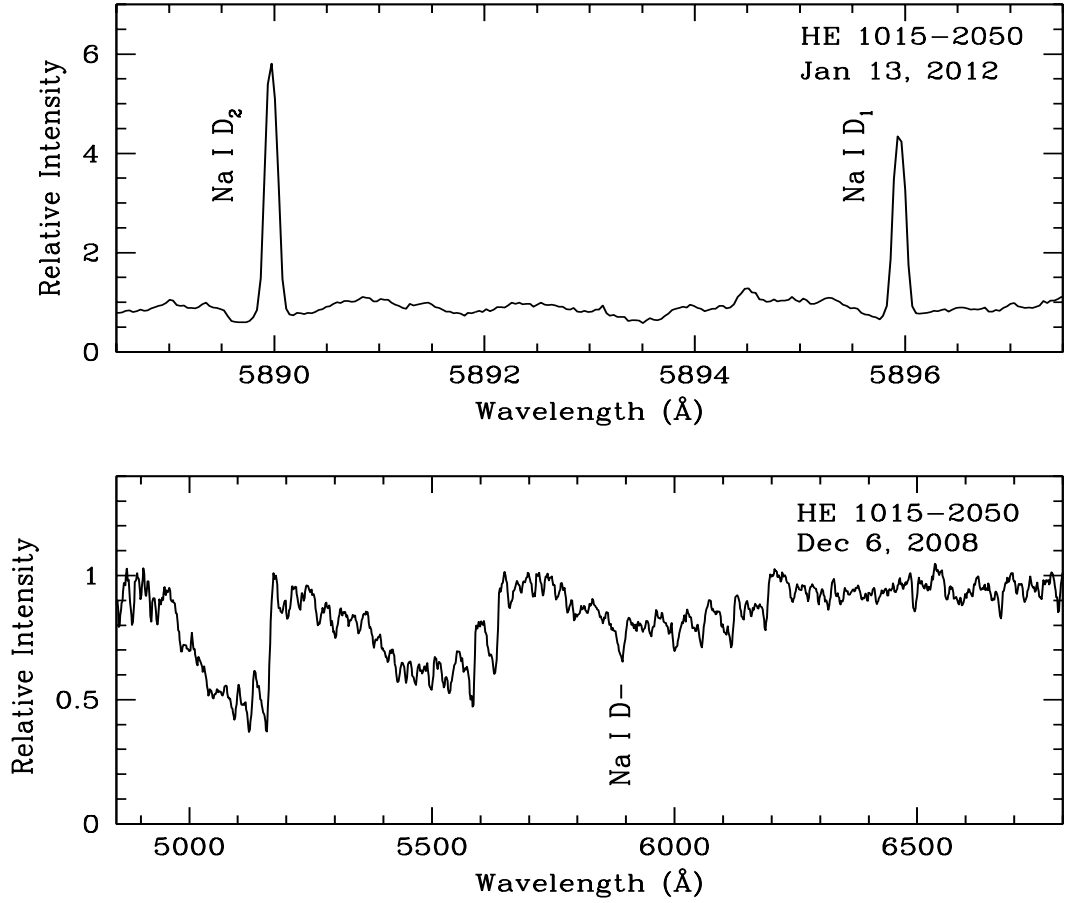}
\caption{ 
Na I D$_{2}$ and Na I D$_{1}$ that appear as one broad absorption feature
in the low-resolution (R ${\sim}$ 1300) spectrum (lower panel) 
appear  in emissions in the high-resolution (R${\sim}$ 50,000) 
spectrum (upper panel).
\label{fig2}
}
\end{figure*}

\begin{figure*}
\centering
\includegraphics[angle=0,height=13cm,width=15cm]{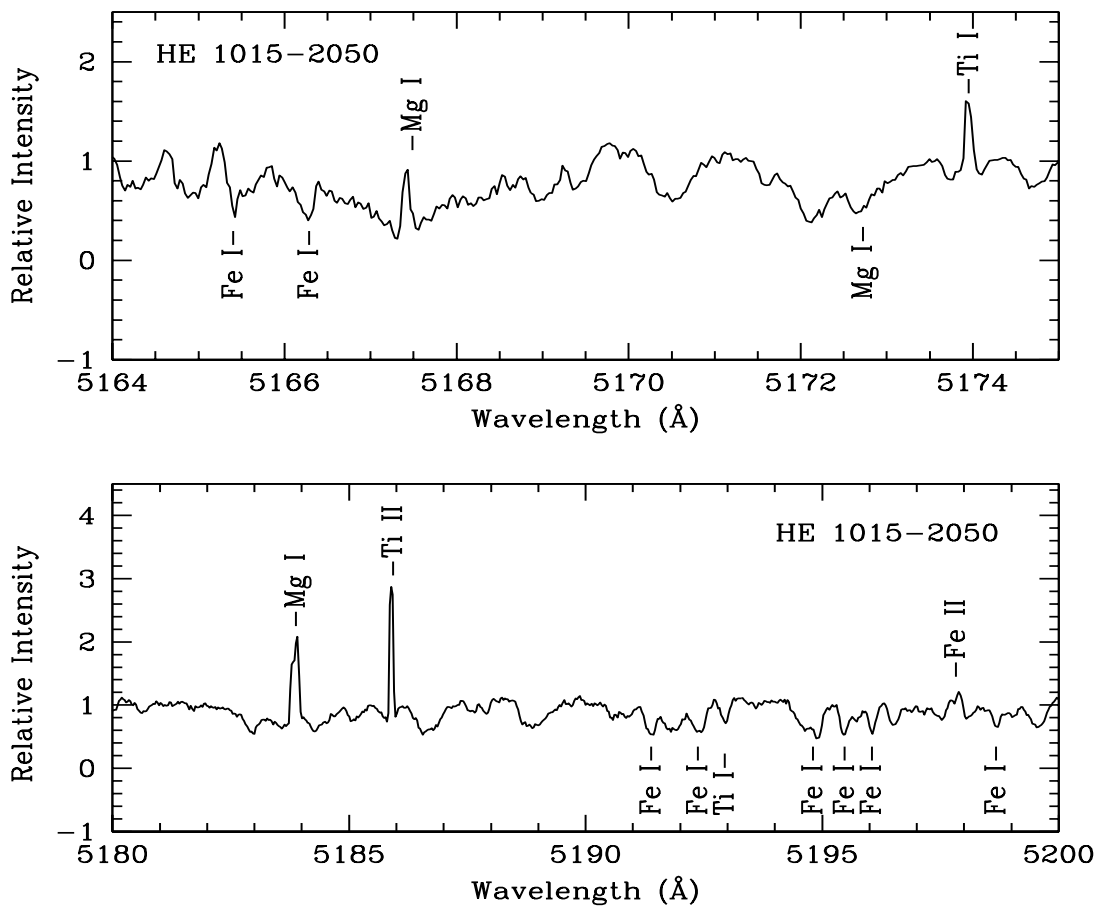}
\caption{ Top:  the  wavelength region 5164-5175 \AA\,
of the spectrum  is shown.
An emission feature at the absorption core of 
Mg I at 5167.3 \AA\,  and Ti I at 5173.7 \AA\, in  emission are indicated
in the figure. 
Two absorption features of Fe I  and the Mg I line at 5172.7 \AA\, are 
 marked.
Bottom:  the wavelength region 5180-5200 \AA\, is shown. Emission
features of Mg I 5183.6 \AA\,, Ti II 5185.9 \AA\,, and Fe II 5197.5 \AA\, 
 are indicated.
Several broad Fe I features and a Ti I feature  seen in absorption are 
marked with vertical lines.  
\label{fig3}
}
\end{figure*}

\begin{figure*}
\centering
\includegraphics[angle=0,height=13cm,width=15cm]{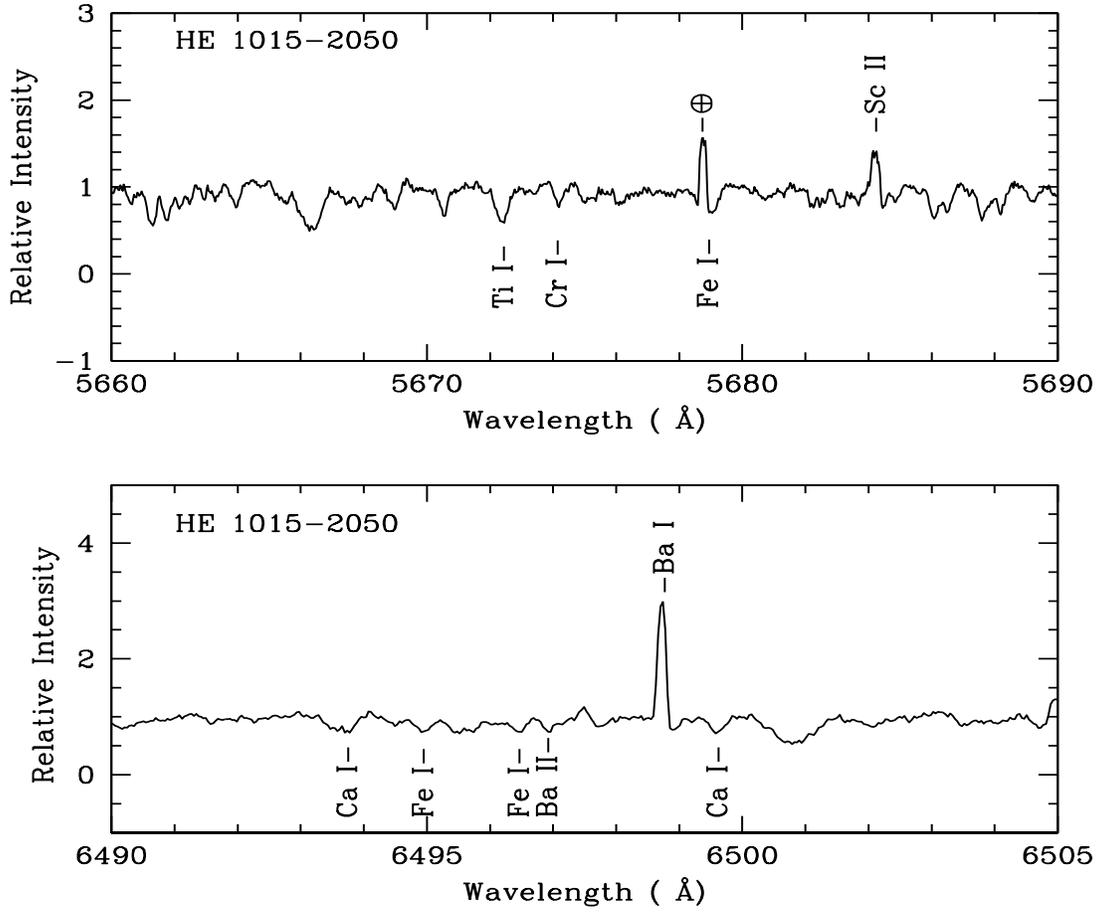}
\caption{ Top: the wavelength region 5660-5690 \AA\, of the spectrum
 is shown. The emission feature of Sc II ${\lambda}$5684.1  and a few  
absorption features are indicated. Bottom: the wavelength region 
6490-6505 \AA\,  of the spectrum is shown. While 
Ba I ${\lambda}$6498.7 is detected in emission ,  Ba II ${\lambda}$6496.9
 is  seen  weakly in  absorption.  Absorption features of Ca I and Fe I
 are also indicated. 
\label{fig4}
}
\end{figure*}

\begin{figure*}
\centering
\includegraphics[angle=0,height=13cm,width=15cm]{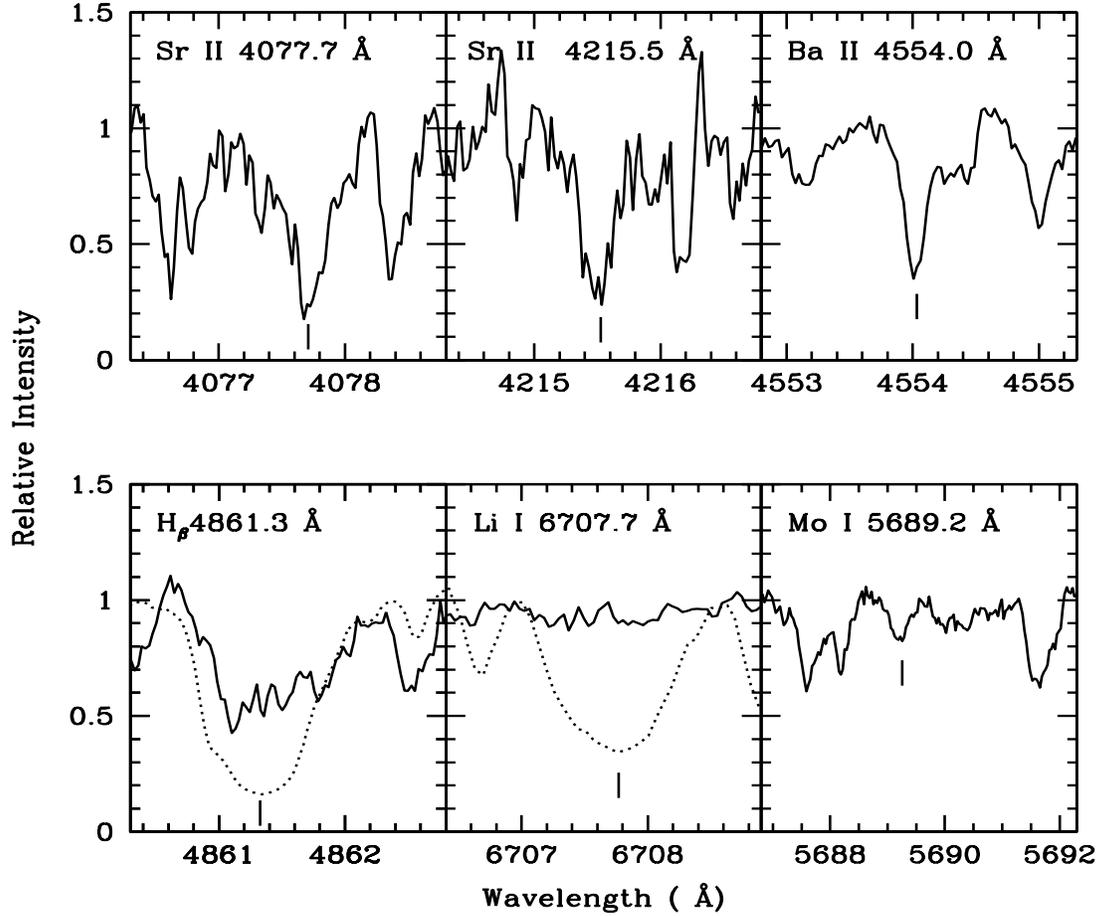}
\caption{ Features due to  H$_{\beta}$, Li,
Mo, Sr, and Ba  are illustrated.  The  vertical
lines indicate the line centers.
 The solid lines  correspond to HE~1015$-$2050 in all the panels.
In the panel showing H$_{\beta}$  the dotted lines that
 correspond to HD~26 and in the panel of Li  the dotted lines
that correspond to Z Umi, a cool RCB star,  are  shown for a comparison.
\label{fig5}
}
\end{figure*}
\end{document}